\documentclass[aps,prx,twocolumn, superscriptaddress]{revtex4-2}  % for review and submission
\usepackage{graphicx}  % needed for figures
\usepackage{dcolumn}   % needed for some tables
\usepackage{bm}        % for math
\usepackage{amssymb}   % for math

\usepackage{physics}
\usepackage{xcolor}
\usepackage[normalem]{ulem}
\usepackage{orcidlink}

\begin{document}

\preprint{APS/123-QED}

\title{Quenching Speculation in Quantum Markets via Entangled Neural Traders}

\author{Kieran Hymas \orcidlink{0000-0003-1761-4298}}
\affiliation{Commonwealth Scientific and Industrial Research Organisation (CSIRO), Clayton, Victoria 3168, Australia}

\author{Hiu Ming Lau}
\affiliation{Commonwealth Scientific and Industrial Research Organisation (CSIRO), Clayton, Victoria 3168, Australia}
\affiliation{RMIT University, Melbourne, Victoria 3001, Australia}

\author{Kareem Raslan}
\affiliation{Commonwealth Scientific and Industrial Research Organisation (CSIRO), Clayton, Victoria 3168, Australia}
\affiliation{School of Electrical and Electronic Engineering, The University of Adelaide, Adelaide, South Australia 5005, Australia}

\author{Qiang Sun}
\affiliation{RMIT University, Melbourne, Victoria 3001, Australia}

\author{Azhar Iqbal}
\affiliation{School of Electrical and Electronic Engineering, The University of Adelaide, Adelaide, South Australia 5005, Australia}

\author{Derek Abbott}
\affiliation{School of Electrical and Electronic Engineering, The University of Adelaide, Adelaide, South Australia 5005, Australia}

\author{Andrew D. Greentree}
\affiliation{RMIT University, Melbourne, Victoria 3001, Australia}

\author{James Q. Quach}
\email{james.quach@csiro.au}
\affiliation{Commonwealth Scientific and Industrial Research Organisation (CSIRO), Clayton, Victoria 3168, Australia}

\date{\today}% It is always \today, today,
             %  but any date may be explicitly specified

\begin{abstract}
Speculative trading can drive pronounced market instabilities, yet existing regulatory and macroprudential tools intervene only after such dynamics emerge. Quantum technologies offer a fundamentally new means of shaping economic behavior by introducing non-classical correlations between decision-makers. Here we demonstrate a prototype quantum stock market in which entanglement between traders’ valuations mitigates the runaway devaluation characteristic of speculative busts. Using reinforcement-learning agents trading a single commodity, we show that replacing classical valuations with quantum-correlated qubit-encoded valuations stabilizes prices and increases the AI traders’ net worth relative to a classical market, where instead agents rapidly converge to liquidation strategies that collapse the asset value. To explain this behavior, we formulate and analyze a quantized version of the $p$-guessing game, a canonical model of speculative dynamics. Quantum entanglement and phase coherence reshape the strategic landscape, eliminating the pathological pure-strategy Nash equilibrium that drives market collapse in the classical game, while mixed-strategy equilibria remain non-degenerate and avoid bust-type outcomes. These results identify quantum correlations as a novel, endogenous mechanism for market stabilization and, more broadly, demonstrate the utility of multi-agent reinforcement learning algorithms for uncovering optimal strategies in complex decision-making frameworks with quantum degrees of freedom.
\end{abstract}

%\keywords{Suggested keywords}%Use showkeys class option if keyword
                              %display desired
\maketitle

%\tableofcontents

\section{Introduction}
Overvaluation from periods of speculative investment can give rise to economic bubbles~\cite{keynes1937general, simsek2021macroeconomics, malpezzi2005role, chang2016review}. Such episodes of rapid and unsustainable expansion are frequently followed by abrupt market corrections, with potentially severe macroeconomic consequences, including recessions, depressions, and political instability~\cite{white2011preventing}. To mitigate or prevent these destabilizing cycles, numerous policy interventions have been proposed, including the sterilization of capital inflows to contain inflationary pressure, macroprudential regulations to limit excessive leverage, and property-tax reforms aimed at curbing speculative behavior in housing markets~\cite{cavoli2006capital, cavoli2017managing, lo2023housing}. The persistence of speculative strategies in market economies has been examined through game-theoretic frameworks such as Keynes’s beauty contest and the more formalized $p$-guessing game~\cite{nagel1995unraveling, duffy1997robustness, ho1998iterated}. These models provide insight into decision-making in uncertain environments, where players react to their perceptions of the intentions of others, leading to the amplification of volatility and the creation of economic bubbles.

The advent of quantum computing and quantum networks presents new opportunities that allow for fundamentally new market behavior, prompting a re-evaluation of economic bubble mitigation strategies. In classical economic models, volatility emerges from the interaction of rational agents exchanging classical information. However, quantum information is conveyed via qubits that behave in a fundamentally different way, with the potential to exist in linear superposition and to be entangled between traders~\cite{nielsen2010quantum}. Consequently, the actions of traders in a quantum stock market are subject to additional quantum correlations, enabling new forms of decision-making and strategic interaction that are not possible in classical environments~\cite{shimamura2004quantum, ikeda2022theory}. In contrast to classical coordination mechanisms, such as regulation, disclosure requirements, or explicit information-sharing protocols, quantum entanglement provides correlations that arise endogenously from the physical medium through which valuations are encoded and exchanged. These correlations reshape the strategic incentives available to traders, constraining destabilizing deviations without requiring centralized oversight or explicit cooperation.

\begin{figure*}[th!]
    \centering
    \includegraphics[width=0.95\linewidth]{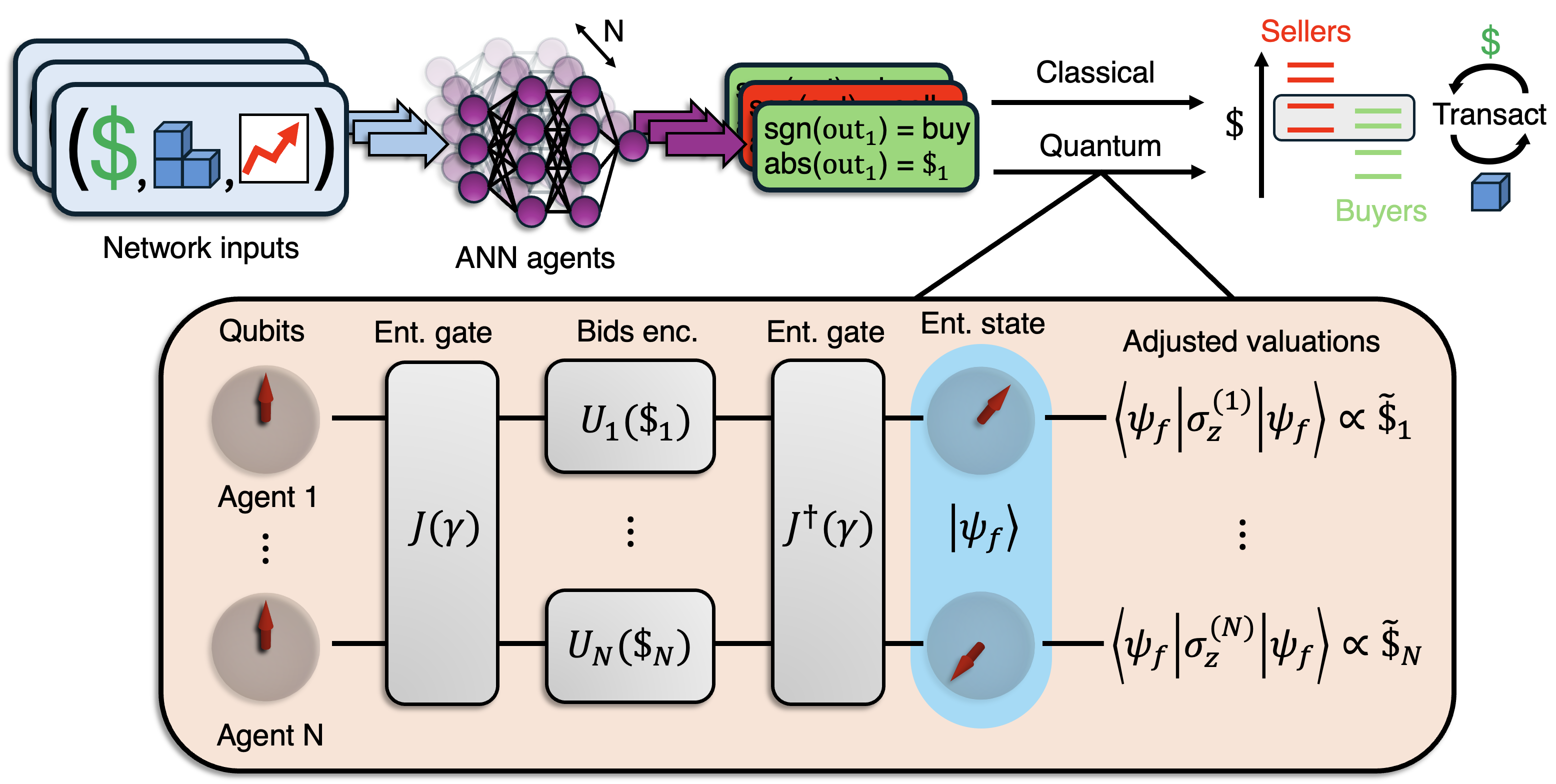}
    \caption{Schematic of the quantum stock market. Each trader is modeled by a neural network that attempts to maximize its net worth by buying and selling a single commodity. At the beginning of each round, the agent’s cash balance, stock holdings, and the previous round’s stock valuation are provided as inputs to its neural network, which enacts a trading policy. The network outputs a real number: its sign determines whether the agent acts as a buyer or seller, and its absolute value specifies the valuation of one unit of stock. In the classical case, buyers and sellers are matched whenever their valuations agree within a given tolerance, and cash is exchanged for a single unit of stock. In the quantum case, before matching, the raw valuations are entangled by a variational quantum circuit (orange box) where they become adjusted before buyers and sellers are matched and transactions occur.}
    \label{fig:1}
\end{figure*}

Here we present a prototypical example of a quantum stock market in which excessive speculation is mitigated. This is achieved by correlating trader’s commodity valuations through the entangled qubits of a quantum computer. Our framework stabilizes the value of the commodity and increases the average net worth of all traders, when compared to analogous simulations in a classical market. Notably, controlled experiments with human participants are often costly and inefficient, and would be further constrained by access to quantum hardware. Training reinforcement learning agents to react optimally to a complex, dynamically evolving environment instead offers a cheaper and more regulated setting to study the consequences of quantum effects on decision-making {\em in silico}. As such, we use policy gradient methods to train a cohort of AI agents to value and trade a single commodity with a goal to maximize their net worth. This sharply distinguishes our work from previous approaches that apply quantum mechanics and quantitative economics merely to produce more accurate representations of classical markets~\cite{piotrowski2002s, piotrowski2003trading, piotrowski2005quantum}.

To explain why speculation is quenched in the quantum market we show that, within the framework of a quantized $p$-guessing game, entangling players’ valuations eliminates the classical game’s pure Nash equilibrium which is responsible for pathological commodity devaluation. Moreover, we perform a computational analysis of the mixed Nash equilibria arising from discretization of the game's strategy space and emphasize that players benefit from a mutual non-zero commodity valuation when they are correlated by entanglement. Although quantum game theory has previously been regarded as a purely mathematical concept with limited practical use, the broader strategic landscapes inherent to quantum games potentially motivate it as an essential framework for decision-making in the era of quantum technology~\cite{wiesner1983conjugate, wiedemann1986quantum, jan2020experimental, chen2022ruling, xu2022experimental, dey2024quantum, perez2024game, khan2025quantum}. More broadly, our approach demonstrates how reinforcement learning agents can serve as a flexible and scalable tool for studying strategic behavior in quantum games, building on the rich quantum game-theoretical literature~\cite{eisert1999quantum, iqbal2001evolutionarily, du2002experimental, flitney2002introduction, piotrowski2003invitation, flitney2009review, ikeda2021infinitely, ghosh2021quantum, frackiewicz2024nash}, and providing a concrete pathway for applying these theoretical frameworks to economically motivated quantum systems.

\section{Trading in a quantum stock market}

\subsection{Reinforcement learning agents}

\begin{figure*}[th!]
    \centering
    \includegraphics[width=0.95\linewidth]{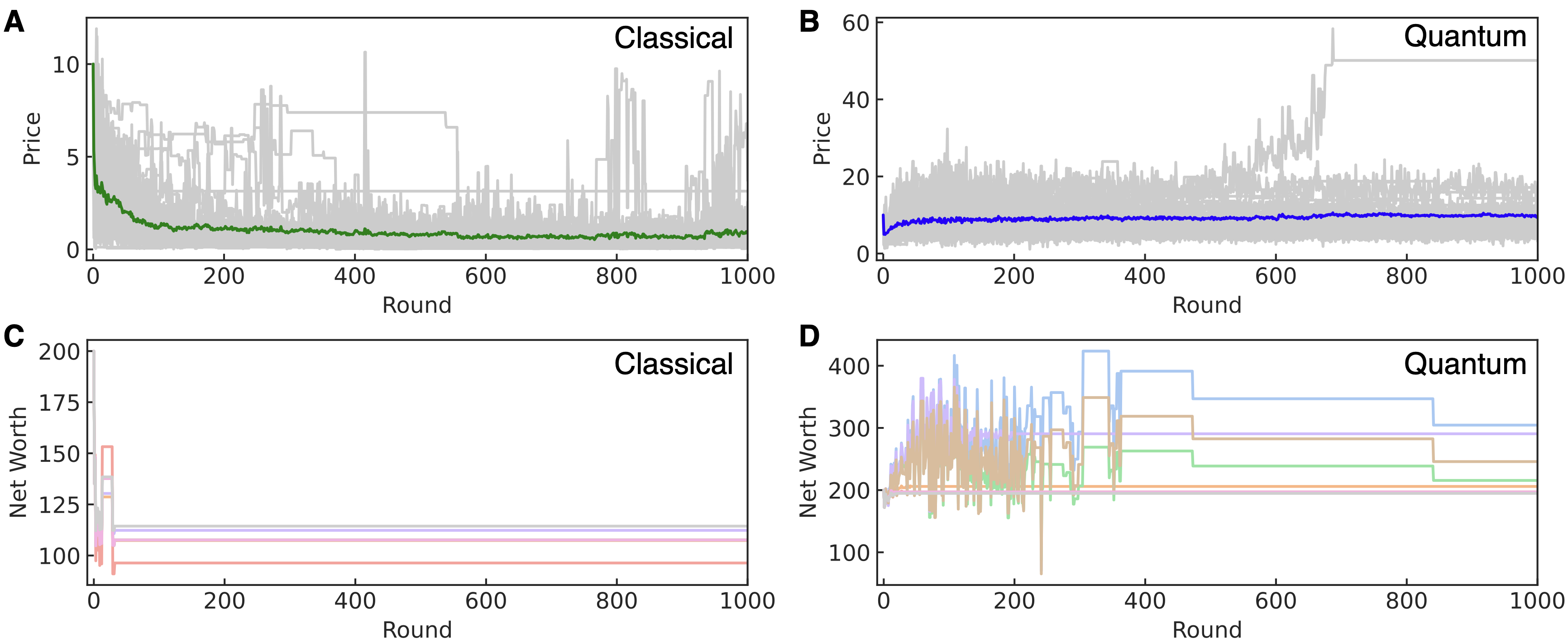}
    \caption{Stock prices and traders’ net worth in the classical and quantum stock market. The price of the commodity in the {\bf A} classical and {\bf B} quantum stock markets, calculated as the average traded price during a given round. The gray curves are commodity valuations from individual runs with random neural network initializations. The colored curves are mean averages of the gray curves over forty runs; the AI traders were randomly initialized at the beginning of each run. The net worth of each of the eight traders (each color corresponds to a different trader) for a single run in the {\bf C} classical and {\bf D} quantum stock markets, at each round of trading. The net worth is calculated for each agent as the current stock holdings multiplied by the current stock valuation plus cash holdings. The net worth of the poorest trader in the quantum market (gray) is greater than the richest trader in the classical market (gray).}
    \label{fig:2}
\end{figure*}

We simulated trading dynamics in a simplified single-commodity stock market with eight traders, where each AI agent’s trading policy was governed by an initially untrained feedforward neural network with one three-dimensional input layer, two 32-dimensional hidden layers with rectified linear activation functions and a one-dimensional output layer. Agents were endowed with initial cash and stock holdings of a commodity with a fixed starting value. The input to the neural network was the agent’s current cash, current stock holdings and the average valuation of the stock in the previous round of trading. The networks output a real number whose sign determines whether the agent intends to buy (positive) or sell (negative) stock at a price $\$_i$ determined by the absolute magnitude of the output. To avoid vanishing gradients associated with sparse positive rewards and to encourage exploration, we added a random number sampled from the standard normal distribution to the output of the network.

Analogous to real markets, agents decided whether to buy or sell based on their internal valuation $\$_i$ and transactions occurred when buyers and sellers expressed compatible valuations, forming matched pairs as illustrated in Fig.~\ref{fig:1}. Buyer and seller pairings were made on a first come first serve basis with a bid-ask tolerance of 1 cash unit. The bid-ask average was deducted from the buyer and given to the seller for the exchange of a single unit of stock per transaction. Once a buyer and seller successfully transacted, they were removed from the trading pool for the remainder of the round.

The neural network weights and biases were updated after each round of trading according to the REINFORCE~\cite{williams1992reinforcement} algorithm using the ADAM~\cite{kingma2015adam} optimizer with learning rate $10^{-3}$. The trainable parameters of the network were updated to implement a policy that maximized the agent’s net worth (current cash $+$ current stock holdings $\times$ current stock market value). At the beginning of each run, the AI agents were reinitialized with random weights and biases, and 10 cash units, 10 stock units. The initial value of the commodity was set to 10 units. 

\subsection{The quantum circuit}

A schematic of the quantum circuit used to entangle traders' valuations is shown in Fig.~\ref{fig:1}. At the beginning of the round, each player's qubit was initialized in the $\ket{0}$ ground state and was subsequently entangled with all other qubits via the entanglement operator
\begin{equation}
J(\gamma) = e^{-i \frac{\gamma}{2} \sigma_x^{(1)} \otimes \dots \otimes \sigma_x^{(N)}}
\end{equation}
where $\sigma_x^{(i)}$ is the Pauli matrix acting on the $i^{\text{th}}$ qubit and $0 \leq \gamma \leq \pi/2$ determines the extent of qubit entanglement. The raw valuations of each player $\$_i$ were input by applying unitary gates $U_i(\theta_i)$ to each qubit that, when represented on the local basis of qubit $i$, reads
\begin{equation}
U_i(\theta_i) = \left( 
\begin{array}{cc}
e^{i \phi_i} \cos\frac{\theta_i}{2} & e^{-i \psi_i} \sin\frac{\theta_i}{2}\\
-e^{i \psi_i} \sin\frac{\theta_i}{2} & e^{-i \phi_i} \cos\frac{\theta_i}{2}
\end{array}
\right)
\end{equation}
where $0 \leq \phi_i \leq 2 \pi$ and $0 \leq \psi_i \leq 2 \pi$ are qubit phases set independently of the raw valuation. In the neural network simulations, $\$_i$ was first scaled by $\pi/\$_{\text{max}}$  to satisfy $0 \leq \theta_i = \pi \$_i/\$_{\text{max}} \leq \pi$ where $\$_{\text{max}}$ is the maximum valuation of that round. The inverse scaling was applied at the end of the circuit. Furthermore, $\phi_i$ and $\psi_i$ were chosen randomly for each qubit at each round however fixing these values across all qubits for all rounds did not significantly change our results. In Section~\ref{sec:3}, we simplify the game theoretical analysis of this circuit by choosing $\psi_i=0$ corresponding to the unitary gate utilized by Eisert et al. in their seminal paper~\cite{eisert1999quantum}.

It is worthwhile to note that restricting the strategy spaces of players in quantum games can violate the principle of strategic symmetry, where players are permitted to employ certain quantum strategies but not the corresponding counter-strategy~\cite{benjamin2001comment}, as can be achieved by removing the closure property of the considered strategy set under composition. Removing this restriction and allowing players to adopt any strategy from $SU(2)$ can change the game fundamentally~\cite{flitney2007nash}. We do not propose to study the full quantum $p$-guessing game in this work but rather utilize the above-mentioned symmetry breaking mechanism as a resource (while still allowing traders access to the full classical strategy space) to engineer a game without a bust equilibrium. Future work could target a quantum stock market where traders leverage $\phi_i$ and $\psi_i$ as well as $\theta_i$ to impose fully quantum strategies with no classical analogues.
    
Finally, the Hermitian conjugate $J^{\dagger}(\gamma)$ is applied resulting in an N-qubit wavefunction $\ket{\psi_f}$ that depends on the valuations $\$_1, \dots, \$_N$ made by each player. The adjusted valuations are associated to the expectation value of each player's qubit through $\tilde{\$}_i = ( 1 + \bra{\psi_f} \sigma_z^{(i)}\ket{\psi_f}) / 2$ which were rescaled back to $\tilde{\$}_i \times \$_{\text{max}}$.

\subsection{Removing speculation with entanglement}

Reinforcement-learning traders serve as adaptive, boundedly rational agents whose strategies evolve in response to realized payoffs, mirroring how human traders learn from market conditions. Their convergence to the classical bust equilibrium in non-entangled markets reflects economically meaningful strategic reasoning rather than artifacts of the learning architecture. As a result, the networks adapted their valuations and trading strategies over successive rounds in response to the observed behavior of other agents.

The main results of our classical and quantum stock market simulations are shown in Fig.~\ref{fig:2}. We tracked the average price of the commodity being traded (Fig.~\ref{fig:2}A and Fig.~\ref{fig:2}B) and the net worth of each AI trader (Fig.~\ref{fig:2}C and Fig.~\ref{fig:2}D) at each round of the simulation. In the classical market, agents collectively anticipated the devaluation of the commodity and sought to maximize their net worth by rapidly liquidating stock holdings in exchange for cash, which retained an intrinsic and agent-independent value in our simulations. This strategy precipitated a sharp decline in the commodity price (Fig.~\ref{fig:2}A), characteristic of a bearish market, leaving all agents poorer by the end of the simulation (Fig.~\ref{fig:2}C). This sub-optimal behavior is well-understood since it embodies the Nash equilibrium of the $p$-guessing game (discussed in Section~\ref{sec:3}), a game theoretic model formulated to study speculation in the stock market~\cite{chang2016review}. In economic terms, the classical Nash equilibrium reflects a self-reinforcing downward spiral in expectations: if each trader anticipates that others will shade their valuations downward, rational behavior induces all agents to converge towards zero. This mechanism mirrors speculative devaluation and recursive belief dynamics in real markets.

In the fully entangled quantum analogue of this market, the classical information $\$_i$ representing each trader’s valuation is replaced by quantum information, encoded through rotations and entanglement of qubits assigned to individual agents. The quantum stock market also showed an initial devaluation of the commodity price as the AI agents learned to maximize their net worth (see Fig.~\ref{fig:2}B). This bust was rapidly corrected and, on average, stabilized near to its initial price, entirely avoiding a bust scenario. Because the commodity retained its initialized value, traders could hold stock without fearing its rapid devaluation, so that their average net worth remained closer to, and sometimes exceeded, its initial value. Fig.~\ref{fig:2}C and Fig.~\ref{fig:2}D compare the net worth of each of the eight traders in an exemplary run. In that case, the poorest trader in the quantum market (gray) exhibited a higher closing net worth $\approx 200$ cash units, than the richest trader in the classical market (gray) with a closing net worth $\approx 120$ cash units.

\begin{figure}[t]
    \centering
    \includegraphics[width=0.85\linewidth]{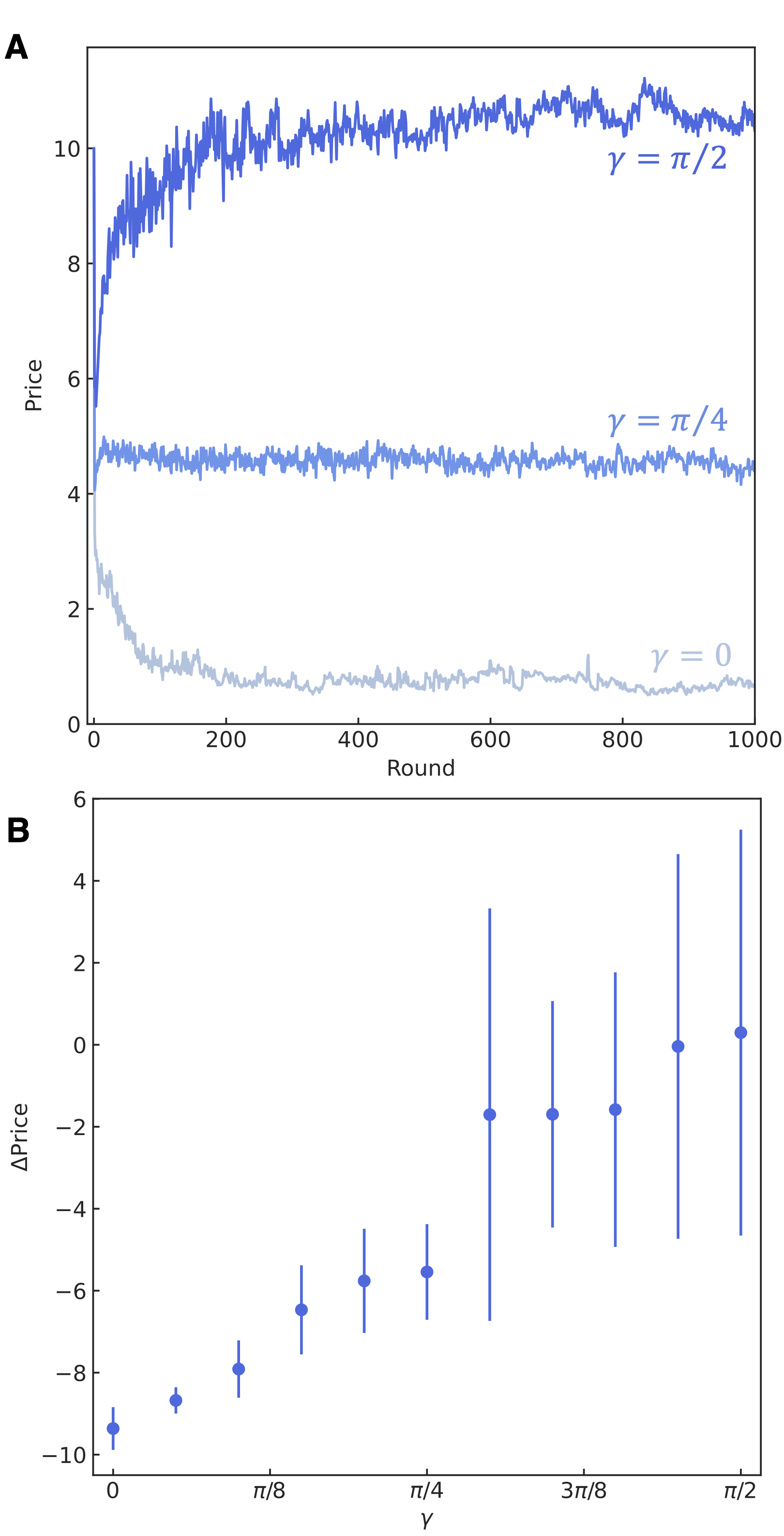}
    \caption{Tuning volatility with entanglement in the quantum stock market. {\bf A} The average commodity price at each round of trading among eight AI agents. The curves are averaged over forty runs in which each AI agent was randomly initialized. The different shadings indicate commodity price for a quantum stock market with no entanglement $\gamma = 0$ (light blue), moderate entanglement $\gamma = \pi / 4$ (blue) and maximal entanglement $\gamma = \pi/2$ (dark blue). {\bf B} Average change in the commodity price $\Delta$Price from a value of 10 cash units after 1,000 rounds of trading in the quantum stock market with different levels of entanglement. The bars indicate one standard deviation of $\Delta$Price.}
    \label{fig:3}
\end{figure}

The degree of entanglement in the quantum circuit provides a fundamentally new lever for controlling volatility in the market. In Fig.~\ref{fig:3} we show the average commodity price over forty trading simulations in the quantum stock market with varying levels of entanglement. When the AI agent’s bids are not entangled ($\gamma = 0$), the quantum stock market functions exactly as the classical market and the commodity price is quickly devalued as in Fig.~\ref{fig:2}A and does not recover. Increasing the degree of entanglement results in varying price stabilization, with maximal entanglement almost returning the stock to its initial intrinsic value. We show the average change in the price of the commodity in Fig.~\ref{fig:3}B at the end of each quantum stock market simulation with varying entanglement, $\gamma$. We observe a monotonic increase in the closing commodity price with $\gamma$, suggesting that the quantum correlations between traders introduced by entanglement could be utilized to shape market dynamics. Price stabilization in the quantum market does not arise from simple averaging of bids. Instead, the entangled circuit transforms the payoff landscape such that extreme undervaluation ceases to be a best response. As a result, reinforcement-learning agents converge to higher-value equilibria because the quantum-mediated environment penalizes strategies that would trigger a classical collapse.

\section{Quantum game theoretic analysis of market speculation} \label{sec:3}

\subsection{Speculation in the classical guessing game}

The $p$-guessing game captures the tendency of rational actors to speculate while trading in bearish and bullish markets, providing a game-theoretic framework to analyze the mechanism underlying market stabilization in the quantum stock market~\cite{nagel1995unraveling, duffy1997robustness, ho1998iterated}. In the $N$-player guessing game each player seeks to maximize their utility
\begin{equation}
u_i(x_1, \dots, x_N) = - \left(x_i - \frac{p}{N} \sum\limits_{j=1}^N x_j \right)^2
\end{equation}
by valuing a commodity $0 \leq x_i \leq 100$ without {\em a priori} knowledge of the valuations made by other players. The value of $p$ determines to what extent agents will under or over value a commodity based on their perception of the average valuation made by all other traders. If $p=1$, the game reduces to guessing the average value of the commodity. When $0<p<1$, a rational player maximizes utility by bidding below the expected mean valuation. If all players adopt this reasoning, the collective valuations converge to zero. This is the only Nash equilibrium~\cite{nash1950equilibrium} of the game and it is quickly learned by the AI agents operating in a classical market (Fig.~\ref{fig:2}A).

\begin{figure*}[t]
    \centering
    \includegraphics[width=0.95\linewidth]{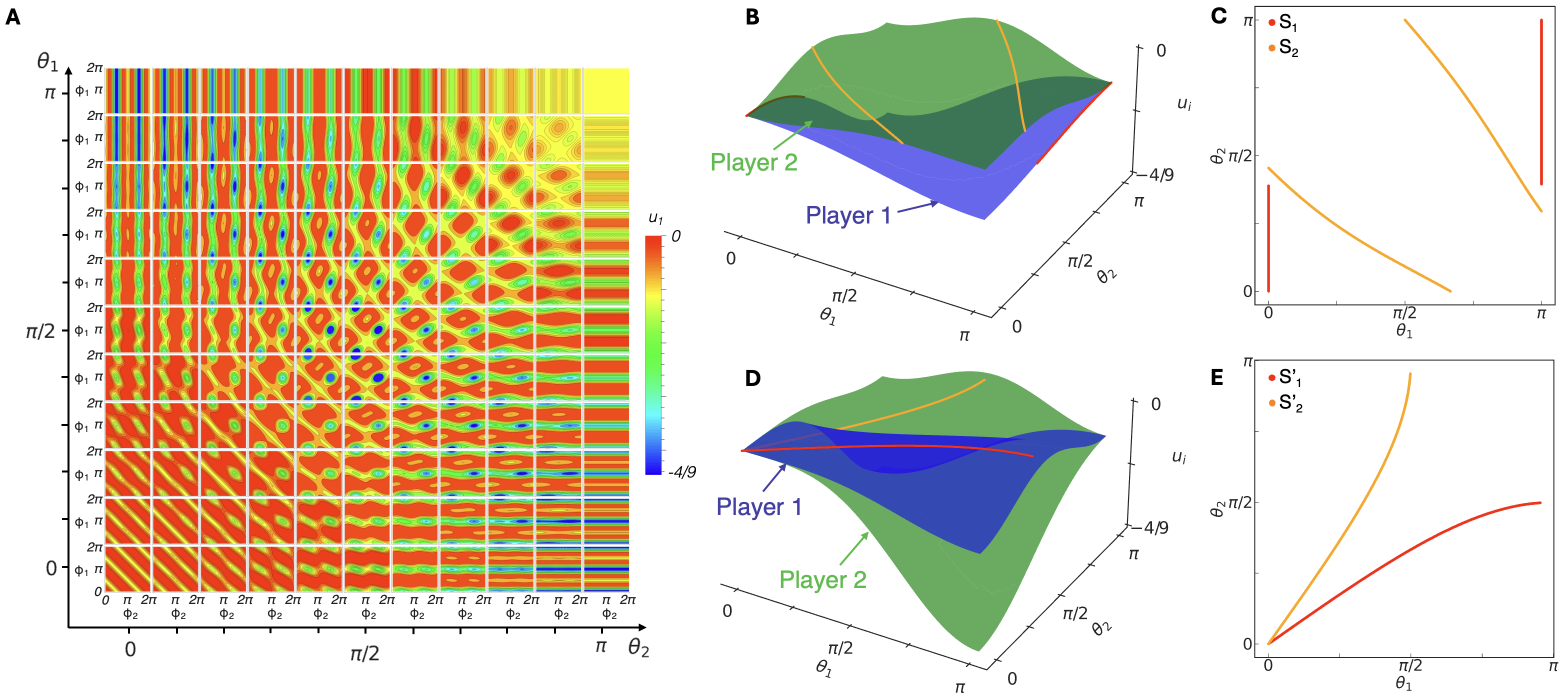}
    \caption{The Nash equilibrium and the two-player quantum guessing game. {\bf A} The utility function of player 1 for the entire strategy space of the two-player $p=\frac{2}{3}$-quantum guessing game when the player’s bids are maximally entangled. {\bf B} The utility functions $u_1$ (blue) and $u_2$ (green) for the maximally entangled two-player $p=\frac{2}{3}$-quantum guessing game as a function of the raw bids $\theta_1$ and $\theta_2$ in the reduced strategy space $(\phi_1, \phi_2 ) = (0, \pi/3)$. The set of optimal responses for Player 1 and Player 2 are represented as red and orange solid lines superimposed to each player's utility function. {\bf C} Each player’s set of optimal responses $S_i$ from b projected onto the $(\theta_1, \theta_2)$-plane. {\bf D} The utility functions $u_1$ (blue) and $u_2$ (green) for the maximally entangled two-player $p=\frac{2}{3}$-quantum guessing game as a function of the raw bids $\theta_1$ and $\theta_2$ in the reduced strategy space $(\phi_1, \phi_2) = (0, 0)$. The set of optimal responses for Player 1 and Player 2 are represented as red and orange solid lines superimposed to each player's utility function. As these sets never intersect, there is no pure Nash equilibrium. {\bf E} Each player’s set of optimal responses $S_i'$ from d projected onto the $(\theta_1, \theta_2)$-plane. The intersection of these sets is a bust Nash equilibrium.}
    \label{fig:4}
\end{figure*}

For the classical $p$-guessing game, the extrema of $u_i (x_1, \dots, x_N)$ are calculated holding all $x_j \neq x_i$ fixed. The solutions $x_i$ correspond to local maxima of player $i$'s utility function since the second derivative $\frac{d^2 u_i}{dx_i^2} = -2(1-p/N)^2 < 0 \ \forall x_1, \dots, x_N$. Thus, a set of optimal responses
\begin{equation}
S_i = \left( x_1, \dots, x_N \right) : x_i = \frac{p}{N-p} \sum\limits_{j \neq i}^N x_j
\end{equation}
can be constructed for each player. If $\mathbf{x}^*=(x_1^*,…,x_N^* ) \in S_1 \cap \dots \cap S_N$, then $\mathbf{x}^*$ corresponds to a strategy in which all players obtain a local maximum of their utility function (i.e. each player does not benefit by unilateral deviation from their valuation) and thus $\mathbf{x}^*$ corresponds to a pure strategy Nash equilibrium~\cite{nash1950equilibrium}. The unique solution to the system of $N$ simultaneous linear equations
\begin{equation}
x_i - \frac{p}{N - P} \sum_{j \neq i} x_j = 0
\end{equation}
with $1 \leq i \leq N$, forms the set of optimal responses and results in a single pure strategy Nash equilibrium $S_1 \cap \dots \cap S_N =(0, \dots, 0)$ for the classical $p$-guessing game. In other words, the only stable equilibrium point of the game occurs when the perceived value of the hypothetical commodity is zero.

\subsection{The quantum $p$-guessing game}

The $p$-guessing game is quantized by identifying $x_i$ in the players’ utility functions $u_i$ with the adjusted valuations $\tilde{\$}_i$ output by the quantum circuit depicted in Fig.~\ref{fig:1}, following the input of the raw valuations $0 \leq \theta_i \leq \pi$. To illustrate how quantum entanglement and qubit phase coherence influence the equilibrium of the quantum game, we focus on the two-player version of the $p=\frac{2}{3}$-guessing game. Fig.~\ref{fig:4}A shows the utility function for Player 1 when both qubits are maximally entangled. The utility landscape acquires a rich dependence on the qubit phase angles $\phi_1$ and $\phi_2$ resembling the feathers of a peacock. In contrast to the classical game, the bust strategy $\left( \theta_1, \phi_1, \theta_2, \phi_2 \right) = \left( 0, \phi_1, 0, \phi_2 \right)$ does not always yield maximal pay-offs; for certain phase configurations (yellow regions in Fig.~\ref{fig:4}A, bottom left), it becomes suboptimal. This additional quantum degree of freedom removes the pure Nash equilibrium corresponding to a total market collapse. Eliminating the classical bust equilibrium fundamentally alters the market’s strategic landscape. Without a stable equilibrium supporting collective undervaluation, downward speculative cascades become strategically dominated. Traders no longer benefit from preemptively liquidating holdings, mitigating the conditions under which bubbles burst or prices collapse.

In Fig.~\ref{fig:4}B we plot the utility functions of each player as a function of their raw bids $\theta_1$ and $\theta_2$ with mismatched phase angles $(\phi_1, \phi_2) = (0, \frac{\pi}{3})$. Superimposed to the utility functions are the sets of optimal responses $S_1$ and $S_2$ each player would adopt if the other player’s strategy was known. These curves are shown explicitly in Fig.~\ref{fig:4}C, projected onto the $(\theta_1, \theta_2)$ plane. The absence of any intersection between these sets demonstrates that no pure-strategy Nash equilibrium exists; for each pair of strategies $(\theta_1, \theta_2)$, at least one player can always improve their outcome by unilateral deviation. The quantum game thus eliminates the pathological bust equilibrium that was adopted by the neural network agents trading in the classical stock market.

Mathematically, the non-trivial extrema of the two-player quantum $p=\frac{2}{3}$-guessing game are calculated as in the classical game by finding the maxima of each player’s utility function with respect to the cosine of the polar angle $\theta_i$ while holding all other player’s strategies fixed. For a given raw valuation $\theta_2$ chosen by Player 2, the extrema of Player 1’s utility occur when
\begin{equation}
\cos\theta_1 = \pm \sqrt{\frac{1-6\cos\theta_2+9\cos^2\theta_2}{13-6\cos\theta_2-3\cos^2\theta_2}}
\end{equation}
where the negative sign applies for $0 \leq \theta_2 < \arccos(1/3)$, and the positive sign for $\arccos(1/3) \leq \theta_2 \leq \pi$. All $(\theta_1,\theta_2)$ combinations except for $(0,0)$ and $(\pi, \pi)$ correspond to local minima in Player 1’s utility function and therefore do not represent optimal responses. The set of optimal responses for Player 1, $S_1$, is instead defined by the boundary conditions  $\theta_1 = 0$ and $\theta_1 = \pi$ as illustrated by the red curves in Fig.~\ref{fig:4}B and Fig.~\ref{fig:4}C. Following a similar analysis, Player 2’s utility function is maximized when
\begin{equation}
\cos\theta_2 = \frac{1 + 4 \cos\theta_1 - 12 \cos^2 \theta_1 \pm 8 \sqrt{6} \Delta (\theta_1)\sin^2\theta_1}{12 \cos^2\theta_1-12 \cos\theta_1 - 49}
\end{equation}
where $\Delta(\theta_1) = \sqrt{5 + 2 \cos \theta_1 - \cos 2\theta_1}$ and the minus and plus signs apply for $0 \leq \theta_1 \leq 2\pi/3$ and $\pi/2 \leq \theta_1 \leq \pi$ respectively. The set of optimal responses $S_2$, is thus constructed from these maxima and are plotted as orange curves in Fig.~\ref{fig:4}B and Fig.~\ref{fig:4}C. That the best response sets share no overlap $S_1 \cap S_2= \emptyset$ follows straightforwardly from substitution and numerical evaluation or from the graphical representation of both sets in Fig.~\ref{fig:4}C.

\subsection{Entanglement is necessary but not sufficient in the two-player game}

For the case with no-phase mismatching $(\phi_1, \phi_2 )=(0,0)$, an analytical solution of the Nash equilibrium is tractable for all values of the entanglement parameter $\gamma$. For this quantum circuit, the adjusted valuation of Player 1 is $(1-\cos^2\gamma \cos\theta_1 - \sin^2\gamma \cos\theta_2)/2$ with a symmetrical output for Player 2: $(1-\cos^2\gamma \cos\theta_2 - \sin^2\gamma \cos\theta_1)/2$. These lead to sets of best responses for Player 1 and Player 2 when
\begin{equation}
\cos\theta_1 = \frac{1 + \left( \cos^2\gamma - 2 \sin^2 \gamma \right) \cos\theta_2}{2 \cos^2\gamma-\sin^2\gamma}
\end{equation}
and
\begin{equation}
\cos\theta_2 = \frac{1 + \left( \cos^2\gamma - 2 \sin^2 \gamma \right) \cos\theta_1}{2 \cos^2\gamma-\sin^2\gamma}
\end{equation}
respectively. Solving these two equations simultaneously reveals a unique intersection at $\cos\theta_1 = \cos\theta_2 = 1$, corresponding to the bust $\theta_1 = \theta_2 = 0$. This result is independent of the entanglement parameter $\gamma$, highlighting that qubit entanglement is a necessary but not a sufficient condition to avoid a bust-type equilibrium point in the two-player game. In Fig.~\ref{fig:4}D and Fig.~\ref{fig:4}E we show this result graphically for maximal entanglement where the sets of best responses $S'_i$ clearly intersect at $(\theta_1, \theta_2) = (0, 0)$, reinstating the bust equilibrium.

\section{Mixed Strategies}

In the absence of pure strategy Nash equilibria, a game may still possess mixed strategy Nash equilibria where players benefit from employing a probabilistic mixture of pure strategies (decided by flipping a biased coin for example)~\cite{osborne2004introduction}. An average strategy $\bar{s}_i = \sum_i w_i s_i$ is the weighted sum of strategies $s_i$ with weights $w_i$ that are determined by the nature of the equilibrium. Finite games always possess at least one mixed-strategy Nash equilibrium, even when no pure-strategy equilibrium exists~\cite{nash1950equilibrium}.

Mixed Nash equilibria calculations were performed using the software package Gambit~\cite{savani2024gambit}. Since Gambit is limited to analyze finite games, we explored a discretized version of the quantum $p=\frac{2}{3}$-guessing game where the strategy space of each player was divided into $k+1$ equally spaced points $\left\{n\pi/k \right\}_{n=0}^k$ allowing us to compute the payoff matrix for each player explicitly. For a given $k$ a bi-matrix game in Gambit was created to solve for all mixed strategy Nash equilibria using the enummixed\_solve function. We studied the mixed strategy equilibria up to a discretization of $k=23$ beyond which the analyses became computationally intractable.

\begin{figure}[t]
    \centering
    \includegraphics[width=0.85\linewidth]{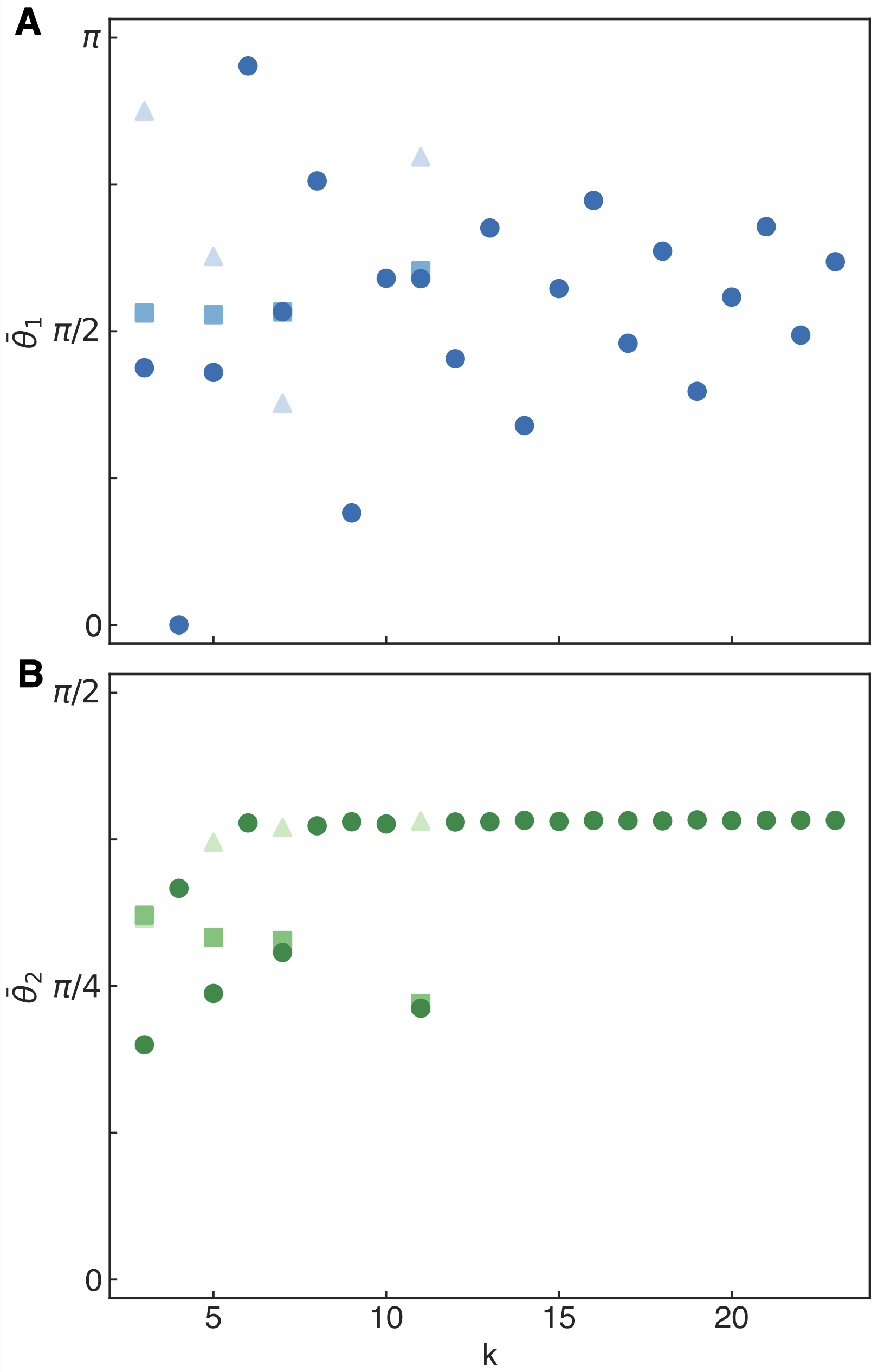}
    \caption{Mixed strategy Nash equilibria of the quantum $p$-guessing game. Average valuation $\bar{\theta}_i$ of {\bf A} Player 1 and {\bf B} Player 2 for the mixed Nash equilibrium (equilibria) of the $k$-discretized quantum $p=\frac{2}{3}$ guessing game when $(\phi_1, \phi_2) = (0, \pi/3)$. When a particular discretization affords more than one mixed Nash equilibrium, the players’ average valuations for each equilibrium point are shown as markers with different shapes, e.g. triangles, squares, circles.}
    \label{fig:5}
\end{figure}

In Fig.~\ref{fig:5} we plot the average valuation $\bar{\theta}_i$ made by each player at a mixed equilibrium point of the game as a function of the discretization parameter $k$. For small discretization (e.g. $k=3$), multiple mixed strategy equilibria exist for the game. For a more fine-grained discretization however (i.e. $11 < k \leq 23$), only one mixed Nash equilibrium persists whereby the average valuations for the players are approximately $\bar{\theta}_1 = \pi / 2$ and $\bar{\theta}_2 = 3 \pi / 8$. Contrary to the classical game which exhibits a pure bust Nash equilibrium, no pure equilibria exist in the restricted quantum game. While the quantum game does tend to a unique mixed strategy equilibrium as the discretization $k$ becomes more fine-grained, the optimal mixed strategies still avoid a bust scenario, as each player values the commodity $\geq \frac{\pi}{2}$, on average.

\section{Discussion and Outlook}
Unlike conventional mechanisms~\cite{bernanke2020newtools, rogoff2017zerobound} designed to curb excessive market speculation such as monetary tightening, capital controls and macroprudential regulation, our approach introduces a fundamentally different lever for market stabilization. By mediating transactions through an entangled quantum circuit, our approach stabilizes markets endogenously through quantum-mediated correlations. Furthermore, our discretized computational analysis shows that busts do not reappear as mixed strategy best responses in the quantum game. This quantum-mechanical mechanism for stabilization is appealing because it arises intrinsically and objectively from the structure of quantum correlations rather than from the intervention of an external agency.

Practically, quantum mediation can be viewed as an extension of the algorithmic intermediaries currently present in modern electronic markets, where order-matching engines already process traders’ bids before execution. A future quantum internet could perform an analogous role, but with quantum correlations enabling stabilization behaviors that have no classical counterpart.
    
We have deliberately focused on a simplified market consisting of a single commodity and agents whose only objective is to maximize their individual net worth, neglecting external influences such as aggregate demand fluctuations, regulatory actions or geopolitical events. Future studies will straightforwardly extend this framework to multi-commodity markets, heterogeneous trader types with differing objectives (e.g., disruptive or risk-averse agents), and dynamically evolving environments. Additionally, our work highlights the utility of AI for studying the outcomes of complex, strategic interactions {\em in silico}, providing a tool that is complementary to quantum game theoretical analysis in lieu of human experimentation. A natural progression of this work would nevertheless see the introduction of human participants into the simulated market to study how quantum mediation influences real decision-making. That human traders would always act as completely rational agents is unlikely~\cite{oaksford2016irrationality, godelier2013rationality}, offering a degree of uncertainty that influences the quantum stock market in a fundamentally new way.

Although our model does not incorporate broader welfare metrics, preventing endogenous price collapse increases traders’ realized net worth and reduces volatility —two properties generally associated with welfare improvements in financial markets. Future extensions could incorporate risk preferences, consumption, or intertemporal utility to evaluate welfare more formally.

Our prototypical quantum stock market assumes that each trader encodes their valuation in a single qubit. Proof-of-principle experiments using real human participants are readily achievable on a single QPU on the scale of typical $p$-guessing game experiments (often involving hundreds of participants). For a practical implementation of the quantum stock market with millions of traders however, this scheme exceeds the number of currently available qubits in even the most advanced quantum computers~\cite{manetsch2025tweezer}. Even so, remarkable progress over the past decade has transformed quantum computing from few-qubit experiments into a rapidly scaling technology~\cite{proctor2025benchmarking}. In parallel, advances in satellite and fiber-based quantum communication, together with improving quantum memories, demonstrate that large-scale quantum networks are becoming technically feasible~\cite{simon2017globalnetwork} so that scaling to a million-qubit quantum stock market may still be achievable through distributed quantum computing. Several promising examples have been shown in recent photonic network prototypes~\cite{aghaeerad2025scaling, main2025distributed} and platforms based on diamond color centers~\cite{chang2025hybrid}. Viewed against this backdrop, the development of a quantum stock market is an increasingly plausible future economic scenario in which market dynamics are shaped by entanglement and other quantum effects.

\section*{Acknowledgments}
K.H. and J.Q.Q. acknowledge funding from the Revolutionary Energy Storage Systems Future Science Platform. K.R., A.I. and D.A. acknowledge support from the Australian Research Council FL240100217 and DP240103404.

\section*{Data Availability}
The computer code for the quantum stock market simulations is available at \url{https://gitlab.com/Ningirama/quantum-stock-market}.

\bibliographystyle{apsrev4-2}
\bibliography{./qsm}

%apsrev4-2.bst 2019-01-14 (MD) hand-edited version of apsrev4-1.bst
%Control: key (0)
%Control: author (72) initials jnrlst
%Control: editor formatted (1) identically to author
%Control: production of article title (-1) disabled
%Control: page (0) single
%Control: year (1) truncated
%Control: production of eprint (0) enabled
\begin{thebibliography}{51}%
\makeatletter
\providecommand \@ifxundefined [1]{%
 \@ifx{#1\undefined}
}%
\providecommand \@ifnum [1]{%
 \ifnum #1\expandafter \@firstoftwo
 \else \expandafter \@secondoftwo
 \fi
}%
\providecommand \@ifx [1]{%
 \ifx #1\expandafter \@firstoftwo
 \else \expandafter \@secondoftwo
 \fi
}%
\providecommand \natexlab [1]{#1}%
\providecommand \enquote  [1]{``#1''}%
\providecommand \bibnamefont  [1]{#1}%
\providecommand \bibfnamefont [1]{#1}%
\providecommand \citenamefont [1]{#1}%
\providecommand \href@noop [0]{\@secondoftwo}%
\providecommand \href [0]{\begingroup \@sanitize@url \@href}%
\providecommand \@href[1]{\@@startlink{#1}\@@href}%
\providecommand \@@href[1]{\endgroup#1\@@endlink}%
\providecommand \@sanitize@url [0]{\catcode `\\12\catcode `\$12\catcode
  `\&12\catcode `\#12\catcode `\^12\catcode `\_12\catcode `\%12\relax}%
\providecommand \@@startlink[1]{}%
\providecommand \@@endlink[0]{}%
\providecommand \url  [0]{\begingroup\@sanitize@url \@url }%
\providecommand \@url [1]{\endgroup\@href {#1}{\urlprefix }}%
\providecommand \urlprefix  [0]{URL }%
\providecommand \Eprint [0]{\href }%
\providecommand \doibase [0]{https://doi.org/}%
\providecommand \selectlanguage [0]{\@gobble}%
\providecommand \bibinfo  [0]{\@secondoftwo}%
\providecommand \bibfield  [0]{\@secondoftwo}%
\providecommand \translation [1]{[#1]}%
\providecommand \BibitemOpen [0]{}%
\providecommand \bibitemStop [0]{}%
\providecommand \bibitemNoStop [0]{.\EOS\space}%
\providecommand \EOS [0]{\spacefactor3000\relax}%
\providecommand \BibitemShut  [1]{\csname bibitem#1\endcsname}%
\let\auto@bib@innerbib\@empty
%</preamble>
\bibitem [{\citenamefont {Keynes}(1937)}]{keynes1937general}%
  \BibitemOpen
  \bibfield  {author} {\bibinfo {author} {\bibfnamefont {J.~M.}\ \bibnamefont
  {Keynes}},\ }\href@noop {} {\bibfield  {journal} {\bibinfo  {journal} {The
  quarterly journal of economics}\ }\textbf {\bibinfo {volume} {51}},\ \bibinfo
  {pages} {209} (\bibinfo {year} {1937})}\BibitemShut {NoStop}%
\bibitem [{\citenamefont {Simsek}(2021)}]{simsek2021macroeconomics}%
  \BibitemOpen
  \bibfield  {author} {\bibinfo {author} {\bibfnamefont {A.}~\bibnamefont
  {Simsek}},\ }\href@noop {} {\bibfield  {journal} {\bibinfo  {journal} {Annual
  Review of Economics}\ }\textbf {\bibinfo {volume} {13}},\ \bibinfo {pages}
  {335} (\bibinfo {year} {2021})}\BibitemShut {NoStop}%
\bibitem [{\citenamefont {Malpezzi}\ and\ \citenamefont
  {Wachter}(2005)}]{malpezzi2005role}%
  \BibitemOpen
  \bibfield  {author} {\bibinfo {author} {\bibfnamefont {S.}~\bibnamefont
  {Malpezzi}}\ and\ \bibinfo {author} {\bibfnamefont {S.}~\bibnamefont
  {Wachter}},\ }\href@noop {} {\bibfield  {journal} {\bibinfo  {journal}
  {Journal of real estate literature}\ }\textbf {\bibinfo {volume} {13}},\
  \bibinfo {pages} {141} (\bibinfo {year} {2005})}\BibitemShut {NoStop}%
\bibitem [{\citenamefont {Chang}\ \emph {et~al.}(2016)\citenamefont {Chang},
  \citenamefont {Newman}, \citenamefont {Walters},\ and\ \citenamefont
  {Wills}}]{chang2016review}%
  \BibitemOpen
  \bibfield  {author} {\bibinfo {author} {\bibfnamefont {V.}~\bibnamefont
  {Chang}}, \bibinfo {author} {\bibfnamefont {R.}~\bibnamefont {Newman}},
  \bibinfo {author} {\bibfnamefont {R.~J.}\ \bibnamefont {Walters}},\ and\
  \bibinfo {author} {\bibfnamefont {G.~B.}\ \bibnamefont {Wills}},\ }\href@noop
  {} {\bibfield  {journal} {\bibinfo  {journal} {International Journal of
  Information Management}\ }\textbf {\bibinfo {volume} {36}},\ \bibinfo {pages}
  {497} (\bibinfo {year} {2016})}\BibitemShut {NoStop}%
\bibitem [{\citenamefont {White}(2011)}]{white2011preventing}%
  \BibitemOpen
  \bibfield  {author} {\bibinfo {author} {\bibfnamefont {L.~J.}\ \bibnamefont
  {White}},\ }\href@noop {} {\bibfield  {journal} {\bibinfo  {journal} {Cato
  J.}\ }\textbf {\bibinfo {volume} {31}},\ \bibinfo {pages} {603} (\bibinfo
  {year} {2011})}\BibitemShut {NoStop}%
\bibitem [{\citenamefont {Cavoli}\ and\ \citenamefont
  {Rajan}(2006)}]{cavoli2006capital}%
  \BibitemOpen
  \bibfield  {author} {\bibinfo {author} {\bibfnamefont {T.}~\bibnamefont
  {Cavoli}}\ and\ \bibinfo {author} {\bibfnamefont {R.~S.}\ \bibnamefont
  {Rajan}},\ }\href@noop {} {\bibfield  {journal} {\bibinfo  {journal} {Asian
  Economic Journal}\ }\textbf {\bibinfo {volume} {20}},\ \bibinfo {pages} {409}
  (\bibinfo {year} {2006})}\BibitemShut {NoStop}%
\bibitem [{\citenamefont {Cavoli}(2017)}]{cavoli2017managing}%
  \BibitemOpen
  \bibfield  {author} {\bibinfo {author} {\bibfnamefont {T.}~\bibnamefont
  {Cavoli}},\ }\href@noop {} {\bibfield  {journal} {\bibinfo  {journal} {Review
  of International Economics}\ }\textbf {\bibinfo {volume} {25}},\ \bibinfo
  {pages} {262} (\bibinfo {year} {2017})}\BibitemShut {NoStop}%
\bibitem [{\citenamefont {Lo}(2023)}]{lo2023housing}%
  \BibitemOpen
  \bibfield  {author} {\bibinfo {author} {\bibfnamefont {H.}~\bibnamefont
  {Lo}},\ }\emph {\bibinfo {title} {The Housing Affordability Crisis: Property
  Tax as a Problem-Solver or Trouble-Maker}},\ \href@noop {} {Ph.D. thesis},\
  \bibinfo  {school} {Harvard University} (\bibinfo {year} {2023})\BibitemShut
  {NoStop}%
\bibitem [{\citenamefont {Nagel}(1995)}]{nagel1995unraveling}%
  \BibitemOpen
  \bibfield  {author} {\bibinfo {author} {\bibfnamefont {R.}~\bibnamefont
  {Nagel}},\ }\href@noop {} {\bibfield  {journal} {\bibinfo  {journal} {The
  American economic review}\ }\textbf {\bibinfo {volume} {85}},\ \bibinfo
  {pages} {1313} (\bibinfo {year} {1995})}\BibitemShut {NoStop}%
\bibitem [{\citenamefont {Duffy}\ and\ \citenamefont
  {Nagel}(1997)}]{duffy1997robustness}%
  \BibitemOpen
  \bibfield  {author} {\bibinfo {author} {\bibfnamefont {J.}~\bibnamefont
  {Duffy}}\ and\ \bibinfo {author} {\bibfnamefont {R.}~\bibnamefont {Nagel}},\
  }\href@noop {} {\bibfield  {journal} {\bibinfo  {journal} {The Economic
  Journal}\ }\textbf {\bibinfo {volume} {107}},\ \bibinfo {pages} {1684}
  (\bibinfo {year} {1997})}\BibitemShut {NoStop}%
\bibitem [{\citenamefont {Ho}\ \emph {et~al.}(1998)\citenamefont {Ho},
  \citenamefont {Camerer},\ and\ \citenamefont {Weigelt}}]{ho1998iterated}%
  \BibitemOpen
  \bibfield  {author} {\bibinfo {author} {\bibfnamefont {T.-H.}\ \bibnamefont
  {Ho}}, \bibinfo {author} {\bibfnamefont {C.}~\bibnamefont {Camerer}},\ and\
  \bibinfo {author} {\bibfnamefont {K.}~\bibnamefont {Weigelt}},\ }\href@noop
  {} {\bibfield  {journal} {\bibinfo  {journal} {The American Economic Review}\
  }\textbf {\bibinfo {volume} {88}},\ \bibinfo {pages} {947} (\bibinfo {year}
  {1998})}\BibitemShut {NoStop}%
\bibitem [{\citenamefont {Nielsen}\ and\ \citenamefont
  {Chuang}(2010)}]{nielsen2010quantum}%
  \BibitemOpen
  \bibfield  {author} {\bibinfo {author} {\bibfnamefont {M.~A.}\ \bibnamefont
  {Nielsen}}\ and\ \bibinfo {author} {\bibfnamefont {I.~L.}\ \bibnamefont
  {Chuang}},\ }\href@noop {} {\emph {\bibinfo {title} {Quantum computation and
  quantum information}}}\ (\bibinfo  {publisher} {Cambridge university press},\
  \bibinfo {year} {2010})\BibitemShut {NoStop}%
\bibitem [{\citenamefont {Shimamura}\ \emph {et~al.}(2004)\citenamefont
  {Shimamura}, \citenamefont {{\"O}zdemir}, \citenamefont {Morikoshi},\ and\
  \citenamefont {Imoto}}]{shimamura2004quantum}%
  \BibitemOpen
  \bibfield  {author} {\bibinfo {author} {\bibfnamefont {J.}~\bibnamefont
  {Shimamura}}, \bibinfo {author} {\bibfnamefont {{\c{S}}.~K.}\ \bibnamefont
  {{\"O}zdemir}}, \bibinfo {author} {\bibfnamefont {F.}~\bibnamefont
  {Morikoshi}},\ and\ \bibinfo {author} {\bibfnamefont {N.}~\bibnamefont
  {Imoto}},\ }\href@noop {} {\bibfield  {journal} {\bibinfo  {journal}
  {International Journal of Quantum Information}\ }\textbf {\bibinfo {volume}
  {2}},\ \bibinfo {pages} {79} (\bibinfo {year} {2004})}\BibitemShut {NoStop}%
\bibitem [{\citenamefont {Ikeda}\ and\ \citenamefont
  {Aoki}(2022)}]{ikeda2022theory}%
  \BibitemOpen
  \bibfield  {author} {\bibinfo {author} {\bibfnamefont {K.}~\bibnamefont
  {Ikeda}}\ and\ \bibinfo {author} {\bibfnamefont {S.}~\bibnamefont {Aoki}},\
  }\href@noop {} {\bibfield  {journal} {\bibinfo  {journal} {Quantum
  Information Processing}\ }\textbf {\bibinfo {volume} {21}},\ \bibinfo {pages}
  {27} (\bibinfo {year} {2022})}\BibitemShut {NoStop}%
\bibitem [{\citenamefont {Piotrowski}(2002)}]{piotrowski2002s}%
  \BibitemOpen
  \bibfield  {author} {\bibinfo {author} {\bibfnamefont {E.}~\bibnamefont
  {Piotrowski}},\ }\href@noop {} {\bibfield  {journal} {\bibinfo  {journal}
  {Quantum market games, Physica A}\ }\textbf {\bibinfo {volume} {312}},\
  \bibinfo {pages} {208} (\bibinfo {year} {2002})}\BibitemShut {NoStop}%
\bibitem [{\citenamefont {Piotrowski}\ and\ \citenamefont
  {S{\l}adkowski}(2003{\natexlab{a}})}]{piotrowski2003trading}%
  \BibitemOpen
  \bibfield  {author} {\bibinfo {author} {\bibfnamefont {E.~W.}\ \bibnamefont
  {Piotrowski}}\ and\ \bibinfo {author} {\bibfnamefont {J.}~\bibnamefont
  {S{\l}adkowski}},\ }\href@noop {} {\bibfield  {journal} {\bibinfo  {journal}
  {International Journal of Theoretical Physics}\ }\textbf {\bibinfo {volume}
  {42}},\ \bibinfo {pages} {1101} (\bibinfo {year}
  {2003}{\natexlab{a}})}\BibitemShut {NoStop}%
\bibitem [{\citenamefont {Piotrowski}\ and\ \citenamefont
  {S{\l}adkowski}(2005)}]{piotrowski2005quantum}%
  \BibitemOpen
  \bibfield  {author} {\bibinfo {author} {\bibfnamefont {E.~W.}\ \bibnamefont
  {Piotrowski}}\ and\ \bibinfo {author} {\bibfnamefont {J.}~\bibnamefont
  {S{\l}adkowski}},\ }\href@noop {} {\bibfield  {journal} {\bibinfo  {journal}
  {Physica A: Statistical Mechanics and its Applications}\ }\textbf {\bibinfo
  {volume} {345}},\ \bibinfo {pages} {185} (\bibinfo {year}
  {2005})}\BibitemShut {NoStop}%
\bibitem [{\citenamefont {Wiesner}(1983)}]{wiesner1983conjugate}%
  \BibitemOpen
  \bibfield  {author} {\bibinfo {author} {\bibfnamefont {S.}~\bibnamefont
  {Wiesner}},\ }\href@noop {} {\bibfield  {journal} {\bibinfo  {journal} {ACM
  Sigact News}\ }\textbf {\bibinfo {volume} {15}},\ \bibinfo {pages} {78}
  (\bibinfo {year} {1983})}\BibitemShut {NoStop}%
\bibitem [{\citenamefont {Wiedemann}(1986)}]{wiedemann1986quantum}%
  \BibitemOpen
  \bibfield  {author} {\bibinfo {author} {\bibfnamefont {D.}~\bibnamefont
  {Wiedemann}},\ }\href@noop {} {\bibfield  {journal} {\bibinfo  {journal} {ACM
  Sigact News}\ }\textbf {\bibinfo {volume} {18}},\ \bibinfo {pages} {48}
  (\bibinfo {year} {1986})}\BibitemShut {NoStop}%
\bibitem [{\citenamefont {Jan}\ \emph {et~al.}(2020)\citenamefont {Jan},
  \citenamefont {Wang}, \citenamefont {Xu}, \citenamefont {Pan}, \citenamefont
  {Chen}, \citenamefont {Han}, \citenamefont {Li}, \citenamefont {Guo},\ and\
  \citenamefont {Abbott}}]{jan2020experimental}%
  \BibitemOpen
  \bibfield  {author} {\bibinfo {author} {\bibfnamefont {M.}~\bibnamefont
  {Jan}}, \bibinfo {author} {\bibfnamefont {Q.-Q.}\ \bibnamefont {Wang}},
  \bibinfo {author} {\bibfnamefont {X.-Y.}\ \bibnamefont {Xu}}, \bibinfo
  {author} {\bibfnamefont {W.-W.}\ \bibnamefont {Pan}}, \bibinfo {author}
  {\bibfnamefont {Z.}~\bibnamefont {Chen}}, \bibinfo {author} {\bibfnamefont
  {Y.-J.}\ \bibnamefont {Han}}, \bibinfo {author} {\bibfnamefont {C.-F.}\
  \bibnamefont {Li}}, \bibinfo {author} {\bibfnamefont {G.-C.}\ \bibnamefont
  {Guo}},\ and\ \bibinfo {author} {\bibfnamefont {D.}~\bibnamefont {Abbott}},\
  }\href@noop {} {\bibfield  {journal} {\bibinfo  {journal} {Advanced Quantum
  Technologies}\ }\textbf {\bibinfo {volume} {3}},\ \bibinfo {pages} {1900127}
  (\bibinfo {year} {2020})}\BibitemShut {NoStop}%
\bibitem [{\citenamefont {Chen}\ \emph {et~al.}(2022)\citenamefont {Chen},
  \citenamefont {Wang}, \citenamefont {Liu}, \citenamefont {Wang},
  \citenamefont {Ying}, \citenamefont {Shang}, \citenamefont {Wu},
  \citenamefont {Gong}, \citenamefont {Deng}, \citenamefont {Liang} \emph
  {et~al.}}]{chen2022ruling}%
  \BibitemOpen
  \bibfield  {author} {\bibinfo {author} {\bibfnamefont {M.-C.}\ \bibnamefont
  {Chen}}, \bibinfo {author} {\bibfnamefont {C.}~\bibnamefont {Wang}}, \bibinfo
  {author} {\bibfnamefont {F.-M.}\ \bibnamefont {Liu}}, \bibinfo {author}
  {\bibfnamefont {J.-W.}\ \bibnamefont {Wang}}, \bibinfo {author}
  {\bibfnamefont {C.}~\bibnamefont {Ying}}, \bibinfo {author} {\bibfnamefont
  {Z.-X.}\ \bibnamefont {Shang}}, \bibinfo {author} {\bibfnamefont
  {Y.}~\bibnamefont {Wu}}, \bibinfo {author} {\bibfnamefont {M.}~\bibnamefont
  {Gong}}, \bibinfo {author} {\bibfnamefont {H.}~\bibnamefont {Deng}}, \bibinfo
  {author} {\bibfnamefont {F.-T.}\ \bibnamefont {Liang}}, \emph {et~al.},\
  }\href@noop {} {\bibfield  {journal} {\bibinfo  {journal} {Physical Review
  Letters}\ }\textbf {\bibinfo {volume} {128}},\ \bibinfo {pages} {040403}
  (\bibinfo {year} {2022})}\BibitemShut {NoStop}%
\bibitem [{\citenamefont {Xu}\ \emph {et~al.}(2022)\citenamefont {Xu},
  \citenamefont {Zhen}, \citenamefont {Yang}, \citenamefont {Cheng},
  \citenamefont {Ren}, \citenamefont {Chen}, \citenamefont {Wang},\ and\
  \citenamefont {Wang}}]{xu2022experimental}%
  \BibitemOpen
  \bibfield  {author} {\bibinfo {author} {\bibfnamefont {J.-M.}\ \bibnamefont
  {Xu}}, \bibinfo {author} {\bibfnamefont {Y.-Z.}\ \bibnamefont {Zhen}},
  \bibinfo {author} {\bibfnamefont {Y.-X.}\ \bibnamefont {Yang}}, \bibinfo
  {author} {\bibfnamefont {Z.-M.}\ \bibnamefont {Cheng}}, \bibinfo {author}
  {\bibfnamefont {Z.-C.}\ \bibnamefont {Ren}}, \bibinfo {author} {\bibfnamefont
  {K.}~\bibnamefont {Chen}}, \bibinfo {author} {\bibfnamefont {X.-L.}\
  \bibnamefont {Wang}},\ and\ \bibinfo {author} {\bibfnamefont {H.-T.}\
  \bibnamefont {Wang}},\ }\href@noop {} {\bibfield  {journal} {\bibinfo
  {journal} {Physical Review Letters}\ }\textbf {\bibinfo {volume} {129}},\
  \bibinfo {pages} {050402} (\bibinfo {year} {2022})}\BibitemShut {NoStop}%
\bibitem [{\citenamefont {Dey}\ \emph {et~al.}(2024)\citenamefont {Dey},
  \citenamefont {Marchetti}, \citenamefont {Caleffi},\ and\ \citenamefont
  {Cacciapuoti}}]{dey2024quantum}%
  \BibitemOpen
  \bibfield  {author} {\bibinfo {author} {\bibfnamefont {I.}~\bibnamefont
  {Dey}}, \bibinfo {author} {\bibfnamefont {N.}~\bibnamefont {Marchetti}},
  \bibinfo {author} {\bibfnamefont {M.}~\bibnamefont {Caleffi}},\ and\ \bibinfo
  {author} {\bibfnamefont {A.~S.}\ \bibnamefont {Cacciapuoti}},\ }\href@noop {}
  {\bibfield  {journal} {\bibinfo  {journal} {IEEE Wireless Communications}\
  }\textbf {\bibinfo {volume} {31}},\ \bibinfo {pages} {90} (\bibinfo {year}
  {2024})}\BibitemShut {NoStop}%
\bibitem [{\citenamefont {P{\'e}rez-Ant{\'o}n}\ \emph
  {et~al.}(2024)\citenamefont {P{\'e}rez-Ant{\'o}n}, \citenamefont
  {L{\'o}pez-S{\'a}nchez},\ and\ \citenamefont {Corbi}}]{perez2024game}%
  \BibitemOpen
  \bibfield  {author} {\bibinfo {author} {\bibfnamefont {R.}~\bibnamefont
  {P{\'e}rez-Ant{\'o}n}}, \bibinfo {author} {\bibfnamefont {J.~I.}\
  \bibnamefont {L{\'o}pez-S{\'a}nchez}},\ and\ \bibinfo {author} {\bibfnamefont
  {A.}~\bibnamefont {Corbi}},\ }\href@noop {} {\bibfield  {journal} {\bibinfo
  {journal} {International Journal of Interactive Multimedia and Artificial
  Intelligence~…}\ } (\bibinfo {year} {2024})}\BibitemShut {NoStop}%
\bibitem [{\citenamefont {Khan}\ \emph {et~al.}(2025)\citenamefont {Khan},
  \citenamefont {Linke}, \citenamefont {Than},\ and\ \citenamefont
  {Baron}}]{khan2025quantum}%
  \BibitemOpen
  \bibfield  {author} {\bibinfo {author} {\bibfnamefont {F.~S.}\ \bibnamefont
  {Khan}}, \bibinfo {author} {\bibfnamefont {N.~M.}\ \bibnamefont {Linke}},
  \bibinfo {author} {\bibfnamefont {A.~T.}\ \bibnamefont {Than}},\ and\
  \bibinfo {author} {\bibfnamefont {D.}~\bibnamefont {Baron}},\ }\href@noop {}
  {\bibfield  {journal} {\bibinfo  {journal} {Quantum Economics and Finance}\
  }\textbf {\bibinfo {volume} {2}},\ \bibinfo {pages} {40} (\bibinfo {year}
  {2025})}\BibitemShut {NoStop}%
\bibitem [{\citenamefont {Eisert}\ \emph {et~al.}(1999)\citenamefont {Eisert},
  \citenamefont {Wilkens},\ and\ \citenamefont
  {Lewenstein}}]{eisert1999quantum}%
  \BibitemOpen
  \bibfield  {author} {\bibinfo {author} {\bibfnamefont {J.}~\bibnamefont
  {Eisert}}, \bibinfo {author} {\bibfnamefont {M.}~\bibnamefont {Wilkens}},\
  and\ \bibinfo {author} {\bibfnamefont {M.}~\bibnamefont {Lewenstein}},\
  }\href@noop {} {\bibfield  {journal} {\bibinfo  {journal} {Physical Review
  Letters}\ }\textbf {\bibinfo {volume} {83}},\ \bibinfo {pages} {3077}
  (\bibinfo {year} {1999})}\BibitemShut {NoStop}%
\bibitem [{\citenamefont {Iqbal}\ and\ \citenamefont
  {Toor}(2001)}]{iqbal2001evolutionarily}%
  \BibitemOpen
  \bibfield  {author} {\bibinfo {author} {\bibfnamefont {A.}~\bibnamefont
  {Iqbal}}\ and\ \bibinfo {author} {\bibfnamefont {A.}~\bibnamefont {Toor}},\
  }\href@noop {} {\bibfield  {journal} {\bibinfo  {journal} {Physics Letters
  A}\ }\textbf {\bibinfo {volume} {280}},\ \bibinfo {pages} {249} (\bibinfo
  {year} {2001})}\BibitemShut {NoStop}%
\bibitem [{\citenamefont {Du}\ \emph {et~al.}(2002)\citenamefont {Du},
  \citenamefont {Li}, \citenamefont {Xu}, \citenamefont {Shi}, \citenamefont
  {Wu}, \citenamefont {Zhou},\ and\ \citenamefont {Han}}]{du2002experimental}%
  \BibitemOpen
  \bibfield  {author} {\bibinfo {author} {\bibfnamefont {J.}~\bibnamefont
  {Du}}, \bibinfo {author} {\bibfnamefont {H.}~\bibnamefont {Li}}, \bibinfo
  {author} {\bibfnamefont {X.}~\bibnamefont {Xu}}, \bibinfo {author}
  {\bibfnamefont {M.}~\bibnamefont {Shi}}, \bibinfo {author} {\bibfnamefont
  {J.}~\bibnamefont {Wu}}, \bibinfo {author} {\bibfnamefont {X.}~\bibnamefont
  {Zhou}},\ and\ \bibinfo {author} {\bibfnamefont {R.}~\bibnamefont {Han}},\
  }\href@noop {} {\bibfield  {journal} {\bibinfo  {journal} {Physical Review
  Letters}\ }\textbf {\bibinfo {volume} {88}},\ \bibinfo {pages} {137902}
  (\bibinfo {year} {2002})}\BibitemShut {NoStop}%
\bibitem [{\citenamefont {Flitney}\ and\ \citenamefont
  {Abbott}(2002)}]{flitney2002introduction}%
  \BibitemOpen
  \bibfield  {author} {\bibinfo {author} {\bibfnamefont {A.~P.}\ \bibnamefont
  {Flitney}}\ and\ \bibinfo {author} {\bibfnamefont {D.}~\bibnamefont
  {Abbott}},\ }\href@noop {} {\bibfield  {journal} {\bibinfo  {journal}
  {Fluctuation and Noise Letters}\ }\textbf {\bibinfo {volume} {2}},\ \bibinfo
  {pages} {R175} (\bibinfo {year} {2002})}\BibitemShut {NoStop}%
\bibitem [{\citenamefont {Piotrowski}\ and\ \citenamefont
  {S{\l}adkowski}(2003{\natexlab{b}})}]{piotrowski2003invitation}%
  \BibitemOpen
  \bibfield  {author} {\bibinfo {author} {\bibfnamefont {E.~W.}\ \bibnamefont
  {Piotrowski}}\ and\ \bibinfo {author} {\bibfnamefont {J.}~\bibnamefont
  {S{\l}adkowski}},\ }\href@noop {} {\bibfield  {journal} {\bibinfo  {journal}
  {International Journal of Theoretical Physics}\ }\textbf {\bibinfo {volume}
  {42}},\ \bibinfo {pages} {1089} (\bibinfo {year}
  {2003}{\natexlab{b}})}\BibitemShut {NoStop}%
\bibitem [{\citenamefont {Flitney}(2009)}]{flitney2009review}%
  \BibitemOpen
  \bibfield  {author} {\bibinfo {author} {\bibfnamefont {A.~P.}\ \bibnamefont
  {Flitney}},\ }\href@noop {} {\bibfield  {journal} {\bibinfo  {journal} {Game
  Theory: Strategies, Equilibria, and Theorems, Nova Science Publishers}\ }
  (\bibinfo {year} {2009})}\BibitemShut {NoStop}%
\bibitem [{\citenamefont {Ikeda}\ and\ \citenamefont
  {Aoki}(2021)}]{ikeda2021infinitely}%
  \BibitemOpen
  \bibfield  {author} {\bibinfo {author} {\bibfnamefont {K.}~\bibnamefont
  {Ikeda}}\ and\ \bibinfo {author} {\bibfnamefont {S.}~\bibnamefont {Aoki}},\
  }\href@noop {} {\bibfield  {journal} {\bibinfo  {journal} {Quantum
  Information Processing}\ }\textbf {\bibinfo {volume} {20}},\ \bibinfo {pages}
  {387} (\bibinfo {year} {2021})}\BibitemShut {NoStop}%
\bibitem [{\citenamefont {Ghosh}(2021)}]{ghosh2021quantum}%
  \BibitemOpen
  \bibfield  {author} {\bibinfo {author} {\bibfnamefont {I.}~\bibnamefont
  {Ghosh}},\ }\href@noop {} {\bibfield  {journal} {\bibinfo  {journal}
  {Resonance}\ }\textbf {\bibinfo {volume} {26}},\ \bibinfo {pages} {671}
  (\bibinfo {year} {2021})}\BibitemShut {NoStop}%
\bibitem [{\citenamefont {Frackiewicz}\ \emph {et~al.}(2024)\citenamefont
  {Frackiewicz}, \citenamefont {Gorczyca-Goraj}, \citenamefont {Grzanka},
  \citenamefont {Nowakowska},\ and\ \citenamefont
  {Szopa}}]{frackiewicz2024nash}%
  \BibitemOpen
  \bibfield  {author} {\bibinfo {author} {\bibfnamefont {P.}~\bibnamefont
  {Frackiewicz}}, \bibinfo {author} {\bibfnamefont {A.}~\bibnamefont
  {Gorczyca-Goraj}}, \bibinfo {author} {\bibfnamefont {K.}~\bibnamefont
  {Grzanka}}, \bibinfo {author} {\bibfnamefont {K.}~\bibnamefont
  {Nowakowska}},\ and\ \bibinfo {author} {\bibfnamefont {M.}~\bibnamefont
  {Szopa}},\ }\href@noop {} {\bibfield  {journal} {\bibinfo  {journal} {arXiv
  preprint arXiv:2411.01711}\ } (\bibinfo {year} {2024})}\BibitemShut {NoStop}%
\bibitem [{\citenamefont {Williams}(1992)}]{williams1992reinforcement}%
  \BibitemOpen
  \bibfield  {author} {\bibinfo {author} {\bibfnamefont {R.~J.}\ \bibnamefont
  {Williams}},\ }\href@noop {} {\bibfield  {journal} {\bibinfo  {journal}
  {Machine Learning}\ }\textbf {\bibinfo {volume} {8}},\ \bibinfo {pages} {229}
  (\bibinfo {year} {1992})}\BibitemShut {NoStop}%
\bibitem [{\citenamefont {Kingma}\ and\ \citenamefont
  {Ba}(2015)}]{kingma2015adam}%
  \BibitemOpen
  \bibfield  {author} {\bibinfo {author} {\bibfnamefont {D.~P.}\ \bibnamefont
  {Kingma}}\ and\ \bibinfo {author} {\bibfnamefont {J.}~\bibnamefont {Ba}},\
  }in\ \href@noop {} {\emph {\bibinfo {booktitle} {Proceedings of the
  International Conference on Learning Representations (ICLR)}}}\ (\bibinfo
  {year} {2015})\BibitemShut {NoStop}%
\bibitem [{\citenamefont {Benjamin}\ and\ \citenamefont
  {Hayden}(2001)}]{benjamin2001comment}%
  \BibitemOpen
  \bibfield  {author} {\bibinfo {author} {\bibfnamefont {S.~C.}\ \bibnamefont
  {Benjamin}}\ and\ \bibinfo {author} {\bibfnamefont {P.}~\bibnamefont
  {Hayden}},\ }\href@noop {} {\bibfield  {journal} {\bibinfo  {journal}
  {Physical Review Letters}\ }\textbf {\bibinfo {volume} {87}},\ \bibinfo
  {pages} {069801} (\bibinfo {year} {2001})}\BibitemShut {NoStop}%
\bibitem [{\citenamefont {Flitney}\ and\ \citenamefont
  {Hollenberg}(2007)}]{flitney2007nash}%
  \BibitemOpen
  \bibfield  {author} {\bibinfo {author} {\bibfnamefont {A.~P.}\ \bibnamefont
  {Flitney}}\ and\ \bibinfo {author} {\bibfnamefont {L.~C.~L.}\ \bibnamefont
  {Hollenberg}},\ }\href@noop {} {\bibfield  {journal} {\bibinfo  {journal}
  {Physics Letters A}\ }\textbf {\bibinfo {volume} {363}},\ \bibinfo {pages}
  {381} (\bibinfo {year} {2007})}\BibitemShut {NoStop}%
\bibitem [{\citenamefont {Nash~Jr}(1950)}]{nash1950equilibrium}%
  \BibitemOpen
  \bibfield  {author} {\bibinfo {author} {\bibfnamefont {J.~F.}\ \bibnamefont
  {Nash~Jr}},\ }\href@noop {} {\bibfield  {journal} {\bibinfo  {journal}
  {Proceedings of the national academy of sciences}\ }\textbf {\bibinfo
  {volume} {36}},\ \bibinfo {pages} {48} (\bibinfo {year} {1950})}\BibitemShut
  {NoStop}%
\bibitem [{\citenamefont {Osborne}\ \emph {et~al.}(2004)\citenamefont {Osborne}
  \emph {et~al.}}]{osborne2004introduction}%
  \BibitemOpen
  \bibfield  {author} {\bibinfo {author} {\bibfnamefont {M.~J.}\ \bibnamefont
  {Osborne}} \emph {et~al.},\ }\href@noop {} {\emph {\bibinfo {title} {An
  introduction to game theory}}},\ Vol.~\bibinfo {volume} {3}\ (\bibinfo
  {publisher} {Springer},\ \bibinfo {year} {2004})\BibitemShut {NoStop}%
\bibitem [{\citenamefont {Savani}\ and\ \citenamefont
  {Turocy}(2024)}]{savani2024gambit}%
  \BibitemOpen
  \bibfield  {author} {\bibinfo {author} {\bibfnamefont {R.}~\bibnamefont
  {Savani}}\ and\ \bibinfo {author} {\bibfnamefont {T.~L.}\ \bibnamefont
  {Turocy}},\ }\href@noop {} {\bibinfo {title} {{Gambit}: The package for
  computation in game theory, version 16.2.0}},\ \bibinfo {howpublished}
  {\url{https://gambit.sourceforge.net/}} (\bibinfo {year} {2024}),\ \bibinfo
  {note} {accessed: 2025-02-03}\BibitemShut {NoStop}%
\bibitem [{\citenamefont {Bernanke}(2020)}]{bernanke2020newtools}%
  \BibitemOpen
  \bibfield  {author} {\bibinfo {author} {\bibfnamefont {B.~S.}\ \bibnamefont
  {Bernanke}},\ }\href@noop {} {\bibfield  {journal} {\bibinfo  {journal}
  {American Economic Review}\ }\textbf {\bibinfo {volume} {110}},\ \bibinfo
  {pages} {943} (\bibinfo {year} {2020})}\BibitemShut {NoStop}%
\bibitem [{\citenamefont {Rogoff}(2017)}]{rogoff2017zerobound}%
  \BibitemOpen
  \bibfield  {author} {\bibinfo {author} {\bibfnamefont {K.}~\bibnamefont
  {Rogoff}},\ }\href@noop {} {\bibfield  {journal} {\bibinfo  {journal}
  {Journal of Economic Perspectives}\ }\textbf {\bibinfo {volume} {31}},\
  \bibinfo {pages} {47} (\bibinfo {year} {2017})}\BibitemShut {NoStop}%
\bibitem [{\citenamefont {Oaksford}\ and\ \citenamefont
  {Hall}(2016)}]{oaksford2016irrationality}%
  \BibitemOpen
  \bibfield  {author} {\bibinfo {author} {\bibfnamefont {M.}~\bibnamefont
  {Oaksford}}\ and\ \bibinfo {author} {\bibfnamefont {S.}~\bibnamefont
  {Hall}},\ }\href@noop {} {\bibfield  {journal} {\bibinfo  {journal} {Trends
  in Cognitive Sciences}\ }\textbf {\bibinfo {volume} {20}},\ \bibinfo {pages}
  {336} (\bibinfo {year} {2016})}\BibitemShut {NoStop}%
\bibitem [{\citenamefont {Godelier}(2013)}]{godelier2013rationality}%
  \BibitemOpen
  \bibfield  {author} {\bibinfo {author} {\bibfnamefont {M.}~\bibnamefont
  {Godelier}},\ }\href@noop {} {\emph {\bibinfo {title} {Rationality and
  Irrationality in Economics}}}\ (\bibinfo  {publisher} {Verso Books},\
  \bibinfo {address} {London},\ \bibinfo {year} {2013})\BibitemShut {NoStop}%
\bibitem [{\citenamefont {Manetsch}\ \emph {et~al.}(2025)\citenamefont
  {Manetsch}, \citenamefont {Nomura}, \citenamefont {Bataille} \emph
  {et~al.}}]{manetsch2025tweezer}%
  \BibitemOpen
  \bibfield  {author} {\bibinfo {author} {\bibfnamefont {H.~J.}\ \bibnamefont
  {Manetsch}}, \bibinfo {author} {\bibfnamefont {G.}~\bibnamefont {Nomura}},
  \bibinfo {author} {\bibfnamefont {E.}~\bibnamefont {Bataille}}, \emph
  {et~al.},\ }\href@noop {} {\bibfield  {journal} {\bibinfo  {journal}
  {Nature}\ }\textbf {\bibinfo {volume} {647}},\ \bibinfo {pages} {60}
  (\bibinfo {year} {2025})}\BibitemShut {NoStop}%
\bibitem [{\citenamefont {Proctor}\ \emph {et~al.}(2025)\citenamefont
  {Proctor}, \citenamefont {Young}, \citenamefont {Baczewski} \emph
  {et~al.}}]{proctor2025benchmarking}%
  \BibitemOpen
  \bibfield  {author} {\bibinfo {author} {\bibfnamefont {T.}~\bibnamefont
  {Proctor}}, \bibinfo {author} {\bibfnamefont {K.}~\bibnamefont {Young}},
  \bibinfo {author} {\bibfnamefont {A.~D.}\ \bibnamefont {Baczewski}}, \emph
  {et~al.},\ }\href@noop {} {\bibfield  {journal} {\bibinfo  {journal} {Nature
  Reviews Physics}\ }\textbf {\bibinfo {volume} {7}},\ \bibinfo {pages} {105}
  (\bibinfo {year} {2025})}\BibitemShut {NoStop}%
\bibitem [{\citenamefont {Simon}(2017)}]{simon2017globalnetwork}%
  \BibitemOpen
  \bibfield  {author} {\bibinfo {author} {\bibfnamefont {C.}~\bibnamefont
  {Simon}},\ }\href@noop {} {\bibfield  {journal} {\bibinfo  {journal} {Nature
  Photonics}\ }\textbf {\bibinfo {volume} {11}},\ \bibinfo {pages} {678}
  (\bibinfo {year} {2017})}\BibitemShut {NoStop}%
\bibitem [{\citenamefont {Aghaee~Rad}\ \emph {et~al.}(2025)\citenamefont
  {Aghaee~Rad}, \citenamefont {Ainsworth}, \citenamefont {Alexander} \emph
  {et~al.}}]{aghaeerad2025scaling}%
  \BibitemOpen
  \bibfield  {author} {\bibinfo {author} {\bibfnamefont {H.}~\bibnamefont
  {Aghaee~Rad}}, \bibinfo {author} {\bibfnamefont {T.}~\bibnamefont
  {Ainsworth}}, \bibinfo {author} {\bibfnamefont {R.~N.}\ \bibnamefont
  {Alexander}}, \emph {et~al.},\ }\href@noop {} {\bibfield  {journal} {\bibinfo
   {journal} {Nature}\ }\textbf {\bibinfo {volume} {638}},\ \bibinfo {pages}
  {912} (\bibinfo {year} {2025})}\BibitemShut {NoStop}%
\bibitem [{\citenamefont {Main}\ \emph {et~al.}(2025)\citenamefont {Main} \emph
  {et~al.}}]{main2025distributed}%
  \BibitemOpen
  \bibfield  {author} {\bibinfo {author} {\bibfnamefont {D.}~\bibnamefont
  {Main}} \emph {et~al.},\ }\href@noop {} {\bibfield  {journal} {\bibinfo
  {journal} {Nature}\ }\textbf {\bibinfo {volume} {638}},\ \bibinfo {pages}
  {383} (\bibinfo {year} {2025})}\BibitemShut {NoStop}%
\bibitem [{\citenamefont {Chang}\ \emph {et~al.}(2025)\citenamefont {Chang},
  \citenamefont {Hou}, \citenamefont {Zhang} \emph {et~al.}}]{chang2025hybrid}%
  \BibitemOpen
  \bibfield  {author} {\bibinfo {author} {\bibfnamefont {X.~Y.}\ \bibnamefont
  {Chang}}, \bibinfo {author} {\bibfnamefont {P.~Y.}\ \bibnamefont {Hou}},
  \bibinfo {author} {\bibfnamefont {W.~G.}\ \bibnamefont {Zhang}}, \emph
  {et~al.},\ }\href@noop {} {\bibfield  {journal} {\bibinfo  {journal} {Nature
  Physics}\ }\textbf {\bibinfo {volume} {21}},\ \bibinfo {pages} {583}
  (\bibinfo {year} {2025})}\BibitemShut {NoStop}%
\end{thebibliography}%

\end{document}